# Publication Trends in Physics Education: A Bibliometric study

Seyedh Mahboobeh Jamali[1]
Ahmad Nurulazam Md Zain[1],
Mohd Ali Samsudin[1]
Nader Ale Ebrahim*[2]

[1]School of Educational Studies (PPIP), Universiti Sains Malaysia, Penang, Malaysia
[2]Centre for Research Services, IPPP, University of Malaya, Malaysia
*Corresponding author: Email: aleebrahim@um.edu.my

**Abstract:**

A publication trend in Physics Education by employing bibliometric analysis leads the researchers to describe current scientific movement. This paper tries to answer "What do Physics education scientists concentrate in their publications?" by analyzing the productivity and development of publications on the subject category of Physics Education in the period 1980–2013. The Web of Science databases in the research areas of "EDUCATION - EDUCATIONAL RESEARCH" was used to extract the publication trends. The study involves 1360 publications, including 840 articles, 503 proceedings paper, 22 reviews, 7 editorial material, 6 Book review, and one Biographical item. Number of publications with "Physical Education" in topic increased from 0.14 % (n = 2) in 1980 to 16.54 % (n = 225) in 2011. Total number of receiving citations is 8071, with approximately citations per papers of 5.93. The results show the publication and citations in Physic Education has increased dramatically while the Malaysian share is well ranked.

*Keywords*: Physics Education, Bibliometrics, Citation analysis, Performance evaluation

## Introduction

Bibliometrics enables researchers to explore the impact of specific field (Uzun, 1996). Bibliometrics is defined as "the study of the quantitative aspects of the production, dissemination, and use of published information" (Moed & Glänzel, 2004, p. 343). Bibliometrics techniques have been used primarily by information scientists to study the growth and distribution of scientific article (Tsai, 2011). Bibliometrics data provided by the monopolist data procedure Web of Science (formerly known as ISI - Institute for Scientific Information), are used for ranking analysis even by non-expert (Weingart, 2005). Eugene Garfield (1979) invention of the ISI is a major breakthrough which enabled statistically analyses of the scientific literature on a very large scale (Moed & Glänzel, 2004). It marks the power of bibliometrics within the studies of science. The bibliometric analysis involves finding trends (Huffman, Baldridge, Bloomfield, Colantonio, Prabhakaran, Ajay, Suh, Lewison, & Prabhakaran, 2013; Menendez-Manjon, Moldenhauer, Wagener, & Barcikowski, 2011; Sooryamoorthy, 2010), correlation and relationship in the author keyword (Chiu & Ho, 2007; Mao, Wang, & Ho, 2010), Keywords Plus® indexed by ISI (Garfield & Sher, 1993; Tan, Fu, & Ho, 2014; Wang, Wang, Zhang, Cai, & Sun, 2013), source titles (Li, Ding, Feng, Wang, & Ho, 2009; Yin, 2013), collaborations with international authors (Jacso, 2012; Zyoud, Al-Jabi, Sweileh, & Awang, 2014), citation analyses (Ho, 2014), language (Diekhoff, Schlattmann, & Dewey, 2013), and many more. Bibliometrics is a specialized and often complex field of study, and far transcends a simple counting of citations (Craig, Plume, McVeigh, Pringle, & Amin, 2007). Citations are now widely accepted as a measurement of recognition. The number of citations that an article received measures its impact on a specific field (Lai, Darius, & Lerut, 2012). Citation analysis is one of the most important tools to evaluate research performance (Bornmann, Schier, Marx, & Daniel, 2012). Citation indicator is important for scientists and universities in all over the world (Farhadi, Salehi, Yunus, Aghaei Chadegani, Farhadi, Fooladi, & Ale Ebrahim, 2013).

Bodin (2012) defines physics education research as research about how we learn, teach, understand, and use physics. Physics Education Research (PER) is the driving force to the way introductory physics is being taught in secondary schools, colleges, and universities (Andre´e Tiberghien, Jossem, & Barojas, 1998; Yeo & Treagust, 2000). Physics education research tends to focus on problems associated with the teaching of physics (Heron & Meltzer, 2005). Physics education research is an interdisciplinary research area and combines education research that is influenced by social studies, psychology, and physics that is a traditional academic subject. Therefore physics education can be approached in many ways depending on the applications (Bodin, 2012). Physic educations included a wide range of studies from research on general culture (Kapitsa, 1982), ''hands-on'' exhibits (Read, 1989), gender issues (Stewart, 1998), classroom-based innovation (Tobias, 2000), multimedia (Wagner, Altherr, Eckert, & Jodl, 2003), IT-based (Akizo, 2004), e-Learning (Stoeva & Cvetkov, 2005), language (Michinel, 2006), used images (Bulbul, 2007), computational problem solving (Landau, 2007), gesture analysis (Scherr, 2008), the use of conceptual diagrams (Martins, Verdeaux, & de Sousa, 2009), quality (Aneta, 2010), 3D Virtual Laboratory (Jeong, Park, Kim, Oh, & Yoo, 2011), modeling (Uhden, Karam, Pietrocola, & Pospiech, 2012), and concept maps (Martinez, Perez, Suero, & Pardo, 2013) to mathematical models (Huang & Fang, 2013).

A major theme at several national and international conferences has been selected on physics education research (McDermott & Redish, 1999). Vollmer (2003) has done a research among 22 European countries experts on physics education research and found that interest in physics education has increased. In particular, the last two decades have seen the growth of an international community engaged in research in physics education (McDermott & Redish, 1999). There are several groups that conduct research in physics education and there is a substantial literature (McDermott, 2001). Most early physics education research focused on student ability to apply the concepts covered in typical introductory university physics courses (Heron & Meltzer, 2005; van Aalst, 2000).

In summary, limited bibliometrics study have been investigated on the publication patterns in the education, such as (Cheng, Wang, Morch, Chen, Kinshuk, & Spector, 2014; Chiang, Kuo, & Yang, 2010; Gomez-Garcia, Ramiro, Ariza, & Granados, 2012; Marshakova-Shaikevich, 2005; Piotrowski, 2013). Additionally, two of them concentrate on physics education (Anduckia, Gomez, & Gomez, 2000; Jacobs & Ingwersen, 2000) which were published more than a decade ago. So, a comprehensive and up-to-date bibliometrics study on Physics education is needed. This paper reports on the used a bibliometric approach to, analyze the productivity and development of publications on the subject category of Physics Education in the period 1980–2013.

## Materials and methods

Most bibliometrics studies have used Web of Science (WoS) to obtain citation data (Bakri & Willett, 2011). Since WoS is the oldest citation database, it has strong coverage with bibliometrics data which goes back to 1900 (Aghaei Chadegani, Salehi, Yunus, Farhadi, Fooladi, Farhadi, & Ale Ebrahim, 2013). The Web of Sciences Core Collection (as a prat of WoS) is a leading database with high quality and multidisciplinary research information, by the subscribed from the Institute of Scientific Information (ISI), also known as Thomson Reuters.

To draw our sample and following prior literature, we first conducted a comprehensive search to collect the data from WoS database. This database provides the information to examine the publication trends in Physics Education since 1980 until 25$^{th}$ December 2013. Science Citation Index Expanded, Social Sciences Citation Index and Arts & Humanities Citation Index, were searched for "Physics Education*" (Physics Education OR Physics Educational and etc.) in the topic (3,770 papers). The search then refined by WoS category "Education Educational Research" (1,360 papers). For each paper, all bibliometric data, especially the number of references and the number of times the paper was cited during the interval between the year of publication and the year 2013, were collected.

In order to find the top cited papers in the field of Physics Education, citations analysis is used. Citation statistics produced by shorter than three years' time frame may not be sufficiently stable (Adams, 2005; UZUN, 2006) because papers appearing in the Web of Science databases over the last few years have not had enough time to accumulate a stable number of citations (Webster, Jonason, & Schember, 2009). Therefore, the time span limited to 1980 until 25 December 2010, yielding a subsample of 813 publications (60% of the original sample). Publications with zero citation were removed. After drawing the final sample, we designed our measurement of the number of citations that studies in our sample had received. The key issue in measuring the citations is that the time elapsed since publication significantly impacts the number of citations that articles receive (Eshraghi, Osman, Gholizadeh, Ali, & Shadgan, 2013). To avoid this bias, we designed and calculated a citation index for each study as the average number of citations per year. Then, we ranked the studies in our sample based on this index to identify top 10 papers with highest citation per year.

In order to answer the question "What do Physics education scientists concentrate in their publications?", the Keyword Plus® and author keywords were extracted from 1360 papers between 1980–2013. Author Keyword are words or phrases provided by the author to justly provide idea of the article while Keywords Plus® are words or phrases that frequently appear in the titles of an article's references, but do not necessarily appear in the title of the article itself. Keywords Plus®, retrieved from the Thomson Reuters auto indexing system, may be present for articles that have no author keywords, or may include important terms not listed among the title, abstract, or author keywords (Ale Ebrahim, 2013).

**Results and discussions**

The aim of conducting bibliometric studies is a statistical analysis of written publications to provide quantitative evaluations. Information produced by bibliometric studies can be exploited as a source to evaluate the performance of sub-fields in a research domain and to adjust science policies with regard to funding allocations and comparing scientific input and output (Debackere & Glanzel, 2004).

*Analysis of author keywords and Keyword Plus®*

The Keywords Plus® analyses for data collected from 1980-2013 were extracted and the percentage range were computed (Table 1). Physics was largely used as the Keywords Plus® (n = 4562) regardless of year category. Among the top 20 Keywords Plus® (n=1726, 37.83% of total), two were used in the titles of the published papers. The words that emerged to be new were the KNOWLEDGE, SCIENCE-EDUCATION, and ACHIEVEMENT. The ACHIEVEMENT, CONCEPTUAL CHANGE, MODEL(S), and INSTRUCTION seemed to be also a keyword that has increased quite a bit. The table indicates that all the importance of Keywords Plus® for future research.

Table 1 Top 20 Keywords Plus®

| NO. | Keyword | Frequency | Percentage of Total |
|---|---|---|---|
| 1 | Physics | 295 | 6.47% |
| 2 | Education | 214 | 4.69% |
| 3 | Science | 196 | 4.30% |
| 4 | Students | 144 | 3.16% |
| 5 | Knowledge | 143 | 3.13% |
| 6 | Science-education | 91 | 1.99% |
| 7 | Achievement | 68 | 1.49% |
| 8 | Conceptual change | 67 | 1.47% |
| 9 | Model(s) | 60 | 1.32% |
| 10 | Instruction | 59 | 1.29% |
| 11 | Classroom | 52 | 1.14% |
| 12 | Conceptions | 48 | 1.05% |
| 13 | Teachers | 42 | 0.92% |
| 14 | Beliefs | 37 | 0.81% |
| 15 | Chemistry | 37 | 0.81% |
| 16 | Misconceptions | 37 | 0.81% |
| 17 | Performance | 37 | 0.81% |
| 18 | Attitudes | 35 | 0.77% |
| 19 | Curriculum | 34 | 0.75% |
| 20 | Views | 30 | 0.66% |

The Author Keywords analysis for data collected is illustrated in Table 2. Top 20 author keywords (589, 14.86% of total) are extracted from 3963 Author Keywords. Consequently, a comparison between Top ten Keywords Plus® and Author Keyword is illustrated in Table 3. There is a considerable difference in the similarity of the Author Keyword and Keywords Plus®. Author Keywords are keywords that are given by the author for each article. On the other hand, the Keywords Plus® is keyword given by the Web of Science for each article. Table 3 lays out the top ten Keywords Plus® and Author Keywords with the frequency in the published article and its rank in the Keywords Plus®. Out of the top ten author keyword only 50% are in the top ten Keywords Plus®. Physics education which is ranked one (n = 295) in the authors keyword fall in n = 118, while Problem solving, Physics education research and Secondary education are not in to 20 Keywords Plus®. The Keywords Plus® looking backward and measure the high frequency phrases in the title of published papers, while the author keyword represent author's current ideas about the paper. So, the order of the authors' keywords which are Physics education, Physics, Science education, Physics education research, Education, Gender, Conceptual change, Misconceptions, and Secondary education are guidelines for the future research topics.

Table 2 Top 20 Keywords

| No. | Keywords | Frequency | Percentage of Total |
|-----|----------|-----------|---------------------|
| 1 | Physics education | 118 | 2.98% |
| 2 | Physics | 85 | 2.14% |
| 3 | Science education | 79 | 1.99% |
| 4 | Physics Education research | 47 | 1.19% |
| 5 | Education | 31 | 0.78% |
| 6 | Gender | 26 | 0.66% |
| 7 | Conceptual change | 25 | 0.63% |
| 8 | Misconceptions | 18 | 0.45% |
| 9 | Secondary education | 18 | 0.45% |
| 10 | Problem solving | 17 | 0.43% |
| 11 | Learning | 14 | 0.35% |
| 12 | Chemistry | 13 | 0.33% |
| 13 | Higher education | 13 | 0.33% |
| 14 | Teacher education | 13 | 0.33% |
| 15 | Teaching | 13 | 0.33% |
| 16 | Technology | 13 | 0.33% |
| 17 | Women in physics | 12 | 0.30% |
| 18 | Constructivism | 12 | 0.30% |
| 19 | Science | 11 | 0.28% |
| 20 | Simulations | 11 | 0.28% |

Table 3 Comparison of Top ten Keywords Plus® and Author Keyword

| Keywords Plus® | N | R | Authors Keyword | N | R |
|---|---|---|---|---|---|
| Physics | 295 | 1 | Physics education | 118 | 1, 2 |
| Education | 214 | 2 | Physics | 85 | 1 |
| Science | 196 | 3 | Science education | 79 | 6 |
| Students | 144 | 4 | Physics education research | 47 | NIL |
| Knowledge | 143 | 5 | Education | 31 | 2 |
| Science-education | 91 | 6 | Gender | 26 | 41 |
| Achievement | 68 | 7 | Conceptual change | 25 | 8 |
| Conceptual change | 67 | 8 | Misconceptions | 18 | 16 |
| Model(s) | 60 | 8 | Secondary education | 18 | NIL |
| Instruction | 59 | 10 | Problem solving | 17 | 87 |

N: Frequency of Keywords Plus®; R: Rank in Keywords Plus®

*Characteristics of publication output*

Over the period of 1980 to 2013, there has been an increase in the number of published paper related to Physics Education, despite the fluctuations seen in Figure 1. In 1980, the numbers of publications were two and the number of publication by 25 December 2013 has risen up to 114. The highest number of publication is in the year of 2011 which was 225. In a wider perspective, the hike in the number of publication begins at 2007, where it leaped 61% from year 2006.

The results were refined by WoS category "Education Educational Research" (1,360 papers, 100%). However, more than 30% of the articles in the physics education research field were published under the Education Scientific Disciplines and Computer Science Interdisciplinary Applications category. The other category ranges from a percentage of 3% - 1% as depicted in Table 4. The categories with lesser than 1% include Computer Science Theory Methods, Management, Computer Science Artificial Intelligence, Engineering Electrical Electronic, Social Sciences Interdisciplinary, Business, Computer Science Software Engineering, Economics, Materials Science Multidisciplinary, and Engineering Mechanical.

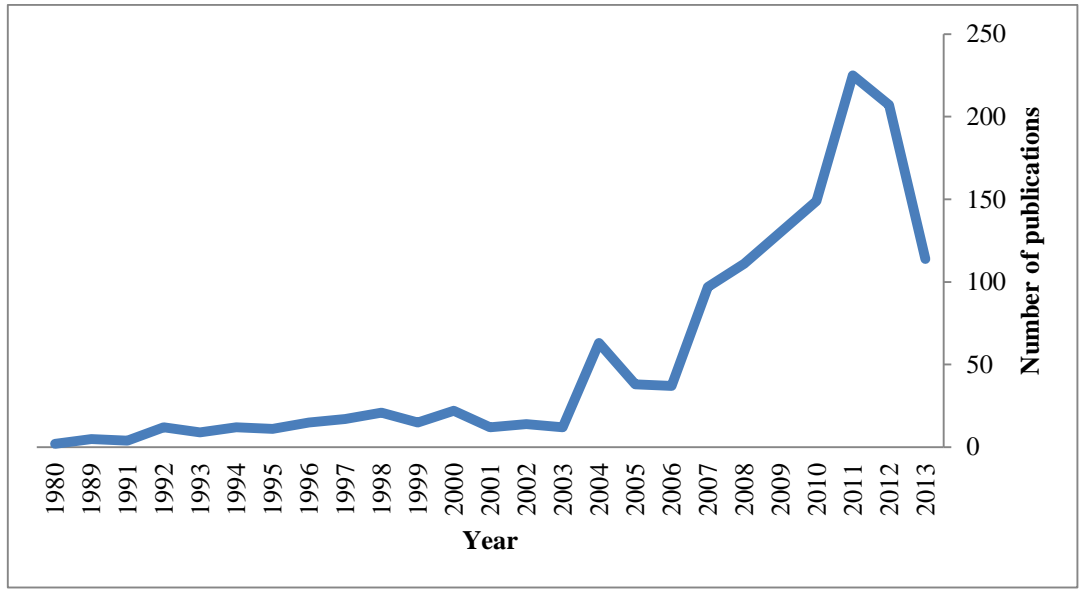

Figure 1 Physics Education publication per year since 1980

Table 4 Web of Science Categories

| No | Web of Science Categories | Records | Percentage of 1360 |
|---|---|---|---|
| 1 | Education educational research | 1360 | 100 |
| 2 | Education scientific disciplines | 318 | 23.3 |
| 3 | Computer science interdisciplinary applications | 107 | 7.8 |
| 4 | Computer science information systems | 42 | 3.0 |
| 5 | History philosophy of science | 39 | 2.8 |
| 6 | Physics applied | 31 | 2.2 |
| 7 | Psychology educational | 28 | 2.0 |
| 8 | Physics multidisciplinary | 27 | 1.9 |
| 9 | Women s studies | 27 | 1.9 |
| 10 | Engineering multidisciplinary | 14 | 1.0 |
| 11 | Information science library science | 14 | 1.0 |

From 1980 to 2013, researchers from top 10 countries published around 70% of total publications in the Physics Education (Figure 2). USA followed by Turkey, Spain and England published around 50% of the publications. Malaysia with 11 publications (0.8%) and Turkey are two developing countries which ranked 2 and 24 respectfully.

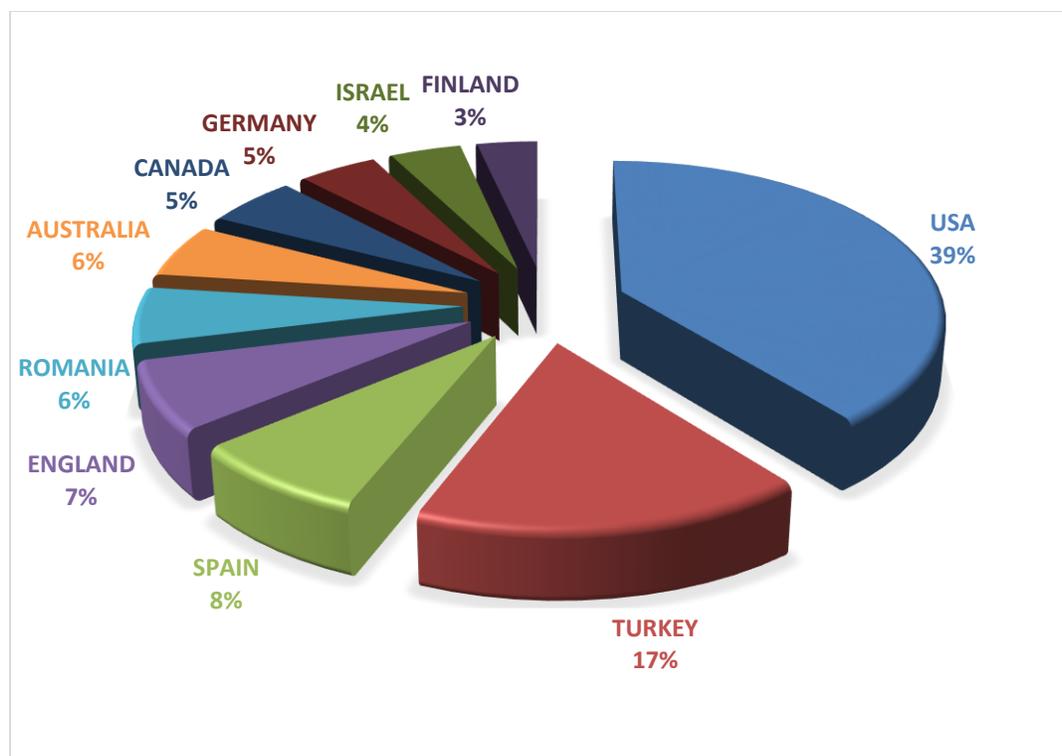

Figure 2 Top 10 active countries on Physics Education publications

*Highly published paper and journal*

As discussed in the methodology the time span for highly cited papers is limited from 1980 to 25 December 2010. Because of, citation statistics produced by shorter than three years' time frame may not be sufficiently stable (Adams, 2005; UZUN, 2006). The journal in which papers are published and the year of publication are investigated to determine the impact on the citation per year. Table 5 shows a list of the top 10 highly cited papers in the Physics Education research. The highest cited article is by (Azar, 2010) with 81 citation per year under the category of Education & Educational Research; Education, Scientific Disciplines, followed by (Kurnaz & Calik, 2009) under the same categories.

Table 5 List of top 10 papers with highest citation per year

| Authors | Title | Citation per year |
| --- | --- | --- |
| Azar, A | A comparison of the effects of two physics laboratory applications with different approaches on student physics achievement | 81 |
| Kurnaz, MA; Calik, M | A thematic review of 'energy' teaching studies: focuses, needs, methods, general knowledge claims and implications | 32 |
| Hazari, Z; Sonnert, G; Sadler, PM; Shanahan, MC | Connecting High School Physics Experiences, Outcome Expectations, Physics Identity, and Physics Career Choice: A Gender Study | 31 |
| Karamustafaoglu, O | Active learning strategies in physics teaching | 27 |
| Trundle, KC; Bell, RL | The use of a computer simulation to promote conceptual change: A quasi-experimental study | 22 |
| Pea, RD | The social and technological dimensions of scaffolding and related theoretical concepts for learning, education, and human activity | 21 |
| Taber, KS; Garcia-Franco, A | Learning Processes in Chemistry: Drawing Upon Cognitive Resources to Learn About the Particulate Structure of Matter | 19 |
| Brewe, E; Sawtelle, V; Kramer, LH; O'Brien, GE; Rodriguez, I; Pamela, P | Toward equity through participation in Modeling Instruction in introductory university physics | 18 |
| Crawford, BA | Learning to teach science as inquiry in the rough and tumble of practice | 17.75 |
| Maltese, AV; Tai, RH | Eyeballs in the Fridge: Sources of early interest in science | 17 |

The top 10 highest published journal and the top 10 highest cited journals in the Physics Education field were analyzed and demonstrated in Table 6. The highest journal that published the Physics Education study is the INTERNATIONAL JOURNAL OF SCIENCE EDUCATION journal with 123 publications from year 1980 – 2010. Among the top 10 journals, 4 of them were existed in the top 10 highest published journal list. The results show that the frequency of publications do not guarantee the quality which reflected by citations per year. The documents type of top cited papers where 11 Article, one Editorial Material and 4 Review papers.

Table 6 Top 10 highly published and highly cited Journals

| **Highly Published journal** | **NoP (n=1360)** | **Rank** | **Highly Cited** | **CPY** | **NoP (n=16)** | **Rank** |
|---|---|---|---|---|---|---|
| International Journal of Science Education | 123 | 1 | Energy Education Science and Technology Part B-Social and Educational Studies | 81, 32, 27.5 | 3 | 34 |
| Journal of Research in Science Teaching | 59 | 2 | Journal of Research in Science Teaching | 31, 17.75 | 2 | 2 |
| Science Education | 43 | 3 | Computers & Education | 22 | 1 | 11 |
| Physical Review Special Topics-Physics Education Research | 40 | 4 | Journal of the Learning Sciences | 21, 19, 14.40 | 3 | 26 |
| Women in Physics | 27 | 5 | Physical Review Special Topics-Physics Education Research | 18 | 2 | 4 |
| Research in Science Education | 23 | 6 | International Journal of Science Education | 17 | 1 | 1 |
| Teaching and Learning of Physics in Cultural Contexts | 21 | 7 | Review of Educational Research | 15.78 | 1 | 54 |
| Innovation and Creativity in Education | 19 | 8 | Journal of Computer Assisted Learning | 15 | 1 | 29 |
| Journal of Science Education and Technology | 15 | 8 | Learning and Instruction | 13.67 | 1 | 40 |
| Teaching and Teacher Education | 8 | 10 | Science Education | 13.22 | 1 | 3 |

*NoP: Number of Publications; CPY: Citation Per Year

## Conclusion

This paper is the first attempt to evaluate the publication trends in physics education as general theme and the research performance, by using bibliometrics methods. 1360 publications from WoS databases were analyzed. The top-ranking authors, articles, and publishing journals were introduced and analyzed, and were ranked by different indexes. The distribution of publications and the trend of publications over the years displayed an increasing trend of publication and revealed the USA as the highest performing country in terms of publishing articles on the subject of Physics Education. In addition, the study prove that the frequency of publications in a journal do not guarantee the quality of the papers which reflected by citations per year. For instance, according to the number of publications journal of "ENERGY EDUCATION SCIENCE AND TECHNOLOGY PART B-SOCIAL AND EDUCATIONAL STUDIES" ranked 34. However, three highly cited papers with 81, 32, 27.5 citations per year, published in the journal. As the research on the Physics Education field grow in interest, bibliometric analysis would assist researchers in identifying key elements and characterization of the Physics Education literature research. In the light of that, more efforts should be channeled in bibliometric studies in all fields.